\documentclass[twocolumn,showpacs,prl]{revtex4}

\usepackage{graphicx}%
\usepackage{dcolumn}
\usepackage{amsmath}

\makeatletter
\def\btt#1{\texttt{\@backslashchar#1}}%
\DeclareRobustCommand\bblash{\btt{\@backslashchar}}%
\makeatother

\begin{document}

\preprint{HEP/123-qed}

\title[Short Title]{Current oscillation and low-field colossal magnetoresistance effect in phase-separated manganites}%

\author{M. Tokunaga}
 \email{mtokunaga@ap.t.u-tokyo.ac.jp}
\author{H. Song}%
\author{Y. Tokunaga}%
\author{T. Tamegai}%
\affiliation{%
Department of Applied Physics, The University of Tokyo, 7-3-1 Hongo, Bunkyo-ku, Tokyo 113-8656, Japan \\
}%

\date{\today}

\begin{abstract}
Current-induced switching from metallic to insulating state is observed in phase-separated states of (La$_{1-y}$Pr$_y$)$_{0.7}$Ca$_{0.3}$MnO$_3$ ($y=0.7$) and Nd$_{0.5}$Ca$_{0.5}$Mn$_{1-z}$Cr$_z$O$_3$ ($z=0.03$) crystals. Application of magnetic fields to this current-induced insulating state causes pronounced low-field negative magnetoresistance effect [${\rm \rho}(H)/{\rm \rho}(0)=10^{-3}$ at $H=1$~kOe]. Application of a constant voltage also causes breakdown of ohmic relation above a threshold voltage. At voltages higher than this threshold value, oscillations in currents are observed. This oscillation is well reproduced by a simple model of local switching of a percolative conduction path.
\end{abstract}

\maketitle

In this decade, manganites with perovskite-type structure have been attracting much attention because of their versatile physical phenomena caused by competing degrees of freedom of charge, spin, lattice, and orbital. Recent extensive studies were first motivated by a discovery of unusual decrease of resistivity by magnetic fields: colossal magnetoresistance (CMR) effects~\cite{Kusters,Helmolt,Chahara,Ju,Jin,Tokura}. Producing the CMR effects at low fields has been a challenging issue.

The large magnetoresistance effect cannot be explained within the framework of the double-exchange (DE) model~\cite{Anderson,deGennes} alone, and needs introduction of additional mechanisms such as electron-phonon coupling mediated by the Jahn-Teller effect~\cite{Millis}. Roles of additional mechanisms become prominent as the DE interaction is suppressed by a reduction of the effective one-electron bandwidth ($W$). In (La,Pr)$_{0.7}$Ca$_{0.3}$MnO$_3$, reduction of $W$ decreases transition temperature to ferromagnetic-metal phase ($T_{\rm MI}$), and makes the CMR effect prominent just above $T_{\rm MI}$~\cite{Hwang}. Recent systematic studies on crystals of $R_{0.55}$Sr$_{0.45}$MnO$_3$ ($R$ is a rare-earth ion)~\cite{Tomioka1} clearly showed the role of short-range correlation of charge/orbital ordering that makes the sample insulating above $T_{\rm MI}$ and hence the CMR effect drastic. Further decrease in $W$ results in long-range charge/orbital ordering. Although magnetic field induced melting of the charge/orbital ordering results in a change in resistivity more than 10 orders of magnitude, the field needed is higher than 10~T~\cite{Tomioka2,Kuwahara,Tokunaga1}, which is beyond the scope of practical applications.

Another aspect of CMR effect is proposed by a scenario of phase separation (PS) into ferromagnetic metal and antiferromagnetic (or paramagnetic) insulator~\cite{Dagotto}. Although it is controversial whether the PS is a necessary condition to cause CMR effects, creation of percolative conduction paths can reproduce drastic change in resistivity~\cite{Mayr}. In some classes of manganites, the presence of co-existing domains with different phases have been directly observed~\cite{Uehara,Fath}. In our previous work, we visualized percolative conduction paths in a phase-separated (La$_{1-y}$Pr$_y$)$_{0.7}$Ca$_{0.3}$MnO$_3$ ($y=0.7$) (LPCMO-0.7) crystal~\cite{Tokunaga2}. We found that percolative conduction in LPCMO-0.7 is collapsed by application of a large amount of current, which coincides with steep increase in resistivity.

In this study, we report anomalous transport properties in the large current/voltage-applied states of LPCMO-0.7: low-field CMR effects in large currents and current oscillations at a constant voltage. To clarify whether these effects are specific to LPCMO-0.7 or not, we carried out similar transport measurements on Cr-doped Nd$_{0.5}$Ca$_{0.5}$MnO$_3$ in which occurrence of PS has been reported~\cite{Kimura}. The results described in this Letter are commonly observed in both samples; therefore, we think it is intrinsic to the PS state.

Crystals of LPCMO-0.7 and Cr-doped Nd$_{0.5}$Ca$_{0.5}$MnO$_3$ were grown by the floating-zone method. Since we need metallic conduction in the ground state, Cr concentration was set to 3\% (Cr3\%-NCMO) so that fraction of metallic phase is larger than the percolation threshold. Details of sample preparation are described in our previous paper~\cite{Tokunaga2}. Sample dimensions are $3100~{\rm \mu m} \times 350~ \mu{\rm m} \times 150~{\rm \mu m}$ for LPCMO-0.7 and $2500~{\rm \mu m} \times 340~ \mu{\rm m} \times 70~{\rm \mu m}$ for Cr3\%-NCMO. Both samples show insulator-metal transition with decreasing temperature below $T_{\rm MI}\sim130$~K [insets of Figs.~1(a) and (b)].

For transport measurements, we utilized several current($I$)/voltage($V$) sources (YOKOGAWA-7651, KEITHLEY-2400, TAKASAGO-BPS40-15). We did not see any instrument dependence for the results shown in this Letter. Magneto-optical (MO) imaging is performed through a polarizing microscope with an indicator film mounted on the sample. Details of MO imaging are described in the previous paper~\cite{Tokunaga2}. Temperature shown in this Letter is the value measured by a thermometer set close to the sample.

Figure~1(a) shows a $V$-$I$ curve of LPCMO-0.7 at 30~K. $V$-$I$ curves shown in this paper were obtained by the standard four-probe method. With increasing $I$, voltage drop in the sample steeply increases by three orders of magnitude at around 13~mA. Above this threshold current ($I_{\rm th}^{\rm up}$), the sample shows negative differential resistance. Successive decrease in $I$ causes a change to the low-resistivity state again at a threshold current ($I_{\rm th}^{\rm down}$) smaller than $I_{\rm th}^{\rm up}$. This $V$-$I$ characteristic is reproduced well, and hence, the whole curve is anti-symmetric against the origin. As reported in our previous paper~\cite{Tokunaga2}, this abrupt increase in resistivity coincides with the change of conduction paths from inhomogeneous to homogeneous.

\begin{figure}[!tb]
\includegraphics[width=7.5cm]{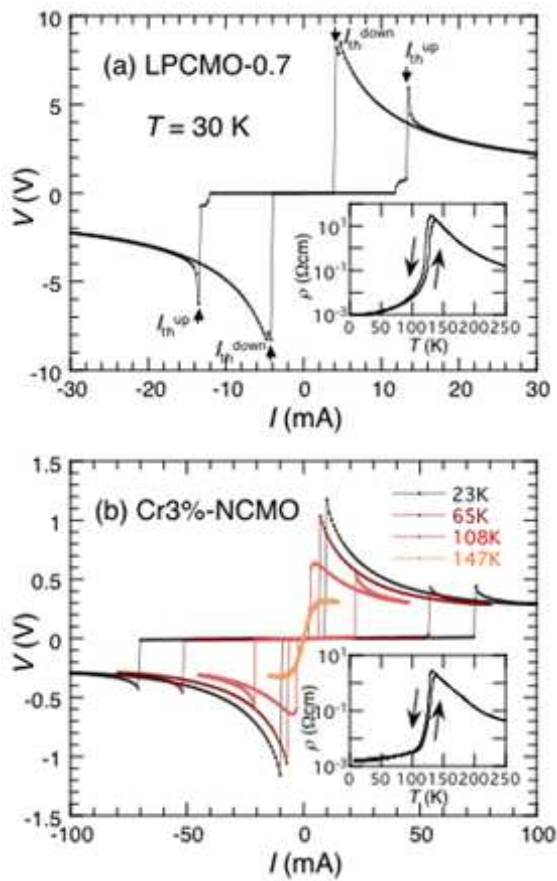}
\caption{(a) A $V$-$I$ curve of LPCMO-0.7 at 30~K. (b) $V$-$I$ curves of Cr3\%-NCMO at various temperatures. It takes about 7~min to obtain one $V-I$ curve. Insets of (a) and (b) show temperature dependence of the resistivity of LPCMO-0.7 and Cr3\%-NCMO measured at $I=10$~$\rm \mu$A.}
\label{fig1}
\end{figure}

Such anomalous $V$-$I$ curves are observed also in Cr3\%-NCMO below $T_{\rm MI}$ as shown in Fig.~1(b). The $I_{\rm th}^{\rm up}$ is 74~mA at 23~K. Although the values of the threshold current density are different between the two samples, it is not clear whether this difference originates from intrinsic nature of these materials or not. By measuring many other pieces of the crystals, we find out that the value of the threshold current density depends not only on the composition, but also on many other factors such as the sample dimensions. Apart from the quantitative differences, qualitative features of $V$-$I$ curves are common in the two systems. Both $I_{\rm th}^{\rm up}$ and $I_{\rm th}^{\rm down}$ decrease monotonically as temperature increases and go to zero at $T_{\rm MI}$.

These $V$-$I$ characteristics are sensitive to magnetic fields ($H$). Figure~2(a) shows the first quadrant of the $V$-$I$ curves of LPCMO-0.7 at various fields. Application of magnetic field increases both $I_{\rm th}^{\rm up}$ and $I_{\rm th}^{\rm down}$. Field dependence of the $I_{\rm th}$ is demonstrated in the inset of Fig.~2(a). This diagram implies that if we apply a current larger than $I_{\rm th}^{\rm up}$ and then reduce it to a certain value slightly higher than $I_{\rm th}^{\rm down}$, we can cause an insulator-metal transition by application of low magnetic field.

\begin{figure}[!tb]
\includegraphics[width=7.5cm]{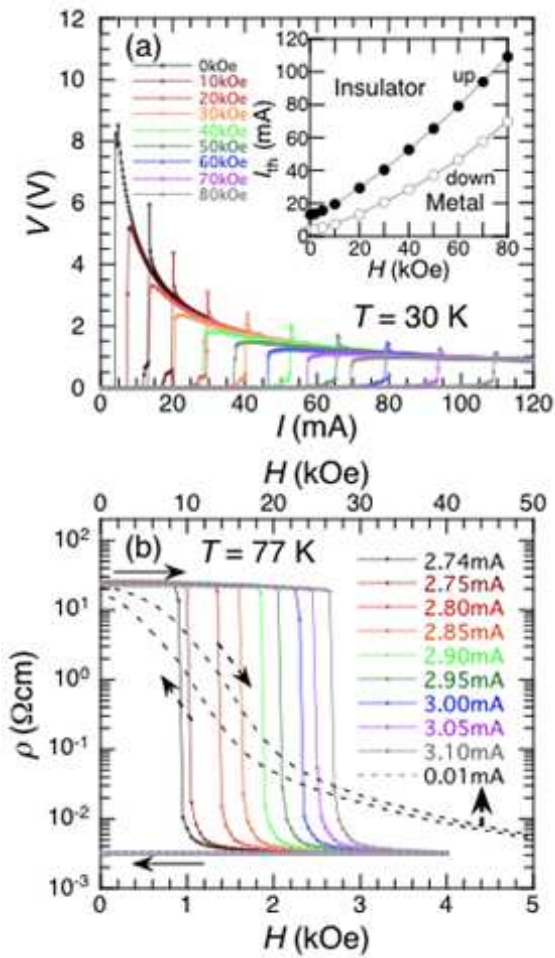}
\caption{(a) Magnetic field dependence of $V-I$ curves of LPCMO-0.7 at 30~K. Inset shows field dependence of $I_{\rm th}^{\rm up}$ and $I_{\rm th}^{\rm down}$. (b) Magnetic field dependence of resistivity at various currents. To make the sample insulating at zero field, a current of 15~mA was applied prior to each measurement. The dotted line indicates the magnetoresistance with $I=10$~$\rm \mu$A at 130~K. Note that the horizontal scales for the dotted line and solid lines are different.}
\label{fig2}
\end{figure}

This low-field CMR effect is indeed observed. Figure~2(b) shows field dependence of resistivity in LPCMO-0.7 at 77~K. In these measurements, we first apply a current of 15~mA to cause a transition to the insulating state, and then reduce it to various values at which we measured magnetoresistance effect. We can see a steep drop of resistivity of 3 orders of magnitude at fields of several kOe. The transition field decreases as the current for measurement decreases toward $I_{\rm th}^{\rm down}$. The sharpness of resistivity drop contrasts with the magnetoresistance effect at low current. The dotted line in Fig.~2(b) represents the magnetoresistance for $I=10$~${\rm \mu}$A (with ten times larger horizontal scale) at 130~K where CMR effect becomes the largest for small currents. Note that resistance at zero field is almost the same in both cases. This fact indicates that although temperature of the environment is measured as 77~K, the sample is locally heated up to about 130~K by application of a current larger than $I_{\rm th}^{\rm up}$. At $I=10$~${\rm \mu}$A, $H=30$~kOe is needed to reduce resistivity down to 10$^{-3}$ of the value at zero field, whereas 0.95~kOe is sufficient at $I=2.74$~mA. Further decrease in current causes spontaneous transition to metallic state without applying external field.

\begin{figure}[!tb]
\includegraphics[width=7.5cm]{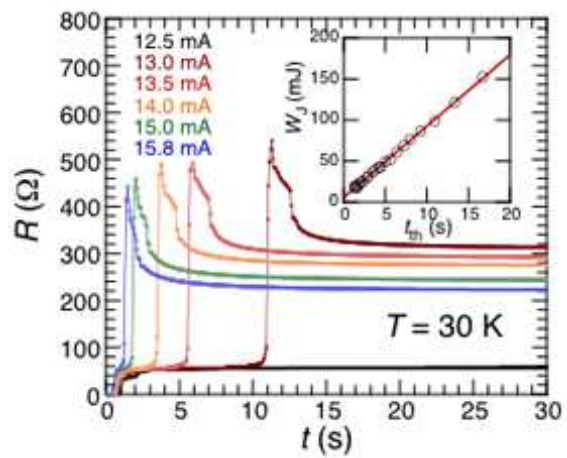}
\caption{Temporal variation of resistivity of LPCMO-0.7 at 30~K after application of various values of currents. Inset shows the relation between total energy put into the sample volume between voltage terminals ($W_{\rm J}$) and time needed to complete the transition ($t_{\rm th}$) (see text).}
\label{fig3}
\end{figure}

This low-field CMR effect is available only after the transition caused by a current larger than $I_{\rm th}^{\rm up}$. This transition is not governed by the magnitude of the current, but by the energy applied to the sample. This is clear in the results of temporal variation of resistivity at various currents (Fig.~3). With increasing $I$, less time is needed to cause transition ($t_{\rm th}$). Let us evaluate the energy spent to cause the transition. Taking into account thermal conductance ($K$) to a thermal bath with temperature $T_{\rm b}$, we can write the net energy needed to cause the transition ($W_{\rm th}$) as
$W_{\rm th}=\int_{0}^{t_{\rm th}}{[VI-K(T-T_{\rm b})]}{\rm d}t.$
We can evaluate the energy of Joule heating $W_{\rm J}=\int_{0}^{t_{\rm th}}{VI}{\rm d}t$ from experimental data. The cooling term, however, is difficult to estimate because of the unknown temperature dependence of $K$. Inset of Fig.~3 shows relation between $W_{\rm J}$ and $t_{\rm th}$. As seen in this figure, $W_{\rm J}$ linearly increases as $t_{\rm th}$ increases. This result means that we can approximate the cooling term $K(T-T_{\rm b})$ by a constant value $p_{\rm c}$, i.e., $W_{\rm J}=W_{\rm th}+p_{\rm c}t_{\rm th}$. From the intercept and the slope of the inset of Fig.~3, we obtain $W_{\rm th}=6$~mJ and $p_{\rm c}=9$~mW. Both the $p_{\rm c}$ and the $W_{\rm th}$ tend to decrease as the $T_{\rm b}$ increases. According to the specific heat data of a related material~\cite{Kiryukhin}, $W_{\rm th}$ of 6~mJ corresponds to the energy to heat the sample volume between the voltage terminals from 30~K to 140~K, which is higher than $T_{\rm MI}$.

\begin{figure}[!tb]
\includegraphics[width=7.5cm]{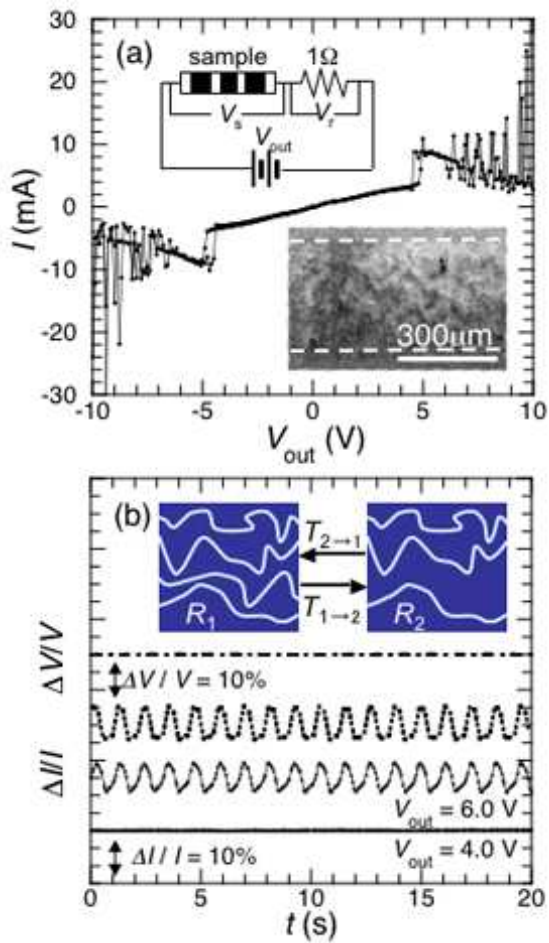}
\caption{(a) An $I-V$ curve of LPCMO-0.7 taken at 30~K. The experimental configuration for this measurement is schematically drawn in the upper inset. The lower inset shows a differential MO image between $V_{\rm out}=+6$~V and $-6$~V. White broken lines are drawn to show the location of the sample edges. (b) Time dependence of the current at $V_{\rm out}=4$ and 6~V (solid lines). The dash-dotted line is the measured voltage that is actually applied to the sample ($V_{\rm s}$) for $V_{\rm out}=6$~V. The dotted line represents the result of simulation described in the text.}
\label{fig4}
\end{figure}

Application of a constant voltage manifests different aspects of anomalous conduction in PS states. Figure~4(a) shows an $I-V$ curve of LPCMO-0.7 at 30~K. For measurements at constant voltages, we used an experimental configuration schematically drawn in the inset of Fig.~4(a). The horizontal axis is the voltage applied by the power source ($V_{\rm out}$). The vertical axis is the current evaluated from the voltage drop at the standard resistance of 1~$\rm \Omega$ ($V_{\rm r}$) connected in series to the sample. In this $I$-$V$ curve, the ohmic relation at low voltage breaks down at around 5~V. In the state above 5~V~\cite{contact}, conduction is still inhomogeneous. The lower inset of Fig.~4(a) is a differential MO image at $V_{\rm out}=\pm6$~V. Inhomogeneous distribution of bright and dark area in the MO image indicates inhomogeneous current flow in the sample at this voltage~\cite{Tokunaga2}.

The noisy $I$-$V$ profile above 5~V originates from temporal variation of $I$ at a constant voltage. Solid lines in Fig.~4(b) show currents at $V_{\rm out}=4$ and 6~V as a function of time. Data at $V_{\rm out}=6$~V show a periodic oscillation of $I$ with the frequency of about 1~Hz. One has to be careful here because instability in the power source may cause such an oscillation when the sample shows negative differential resisitance~\cite{Ogawa}. However, this is not the case for the present observation. The dash-dotted line in Fig.~4(b) represents the voltage drop at the sample ($V_{\rm s}$). We cannot see significant change in $V_{\rm s}$. Thereby, observed oscillations of $I$ cannot be ascribed to instability of the voltage source. The current oscillation survives more than 1 hour unless we change the experimental condition. Such current oscillations are also observed in Cr3\%-NCMO. Therefore, we think this phenomenon is intrinsic in PS states. The frequency and amplitude change with changing $V_{\rm out}$, temperature, and magnetic field. Details of the change in oscillation will be reported in a separate paper.

In the following, let us discuss the origin of this current oscillation. Application of large current or voltage causes heating of the sample and closes conduction paths. Let us consider a situation where a certain conduction path closes upon heating caused by an applied voltage $V$ [see insets of Fig.~4(b)]. We define the resistance with and without the path as $R_1$ and $R_2$, respectively ($R_1<R_2$). If the cooling power $p_{\rm c}$ satisfies $V^2/R_2<p_{\rm c}<V^2/R_1$, the sample tends to warm up (cool down) at $R_1$ ($R_2$). Let us assume the discontinuous change between the two states and hysteretic response against temperature change, i.e. annihilation (creation) of the path takes place at $T_{1 \rightarrow 2}$ ($T_{2 \rightarrow 1}$), where $T_{1 \rightarrow 2}>T_{2 \rightarrow 1}$. Introduction of the hysteresis is consistent with temperature dependence of resistivity shown in insets of Figs.~1(a) and (b). The observed oscillation of current at $V_{\rm out}=6$~V corresponds to that of $R$ between 740 and 800~${\rm \Omega}$ at a frequency $f=0.88$~Hz. Taking the heat capacity $C=0.5$~mJ/K~\cite{Kiryukhin}, we obtain this frequency when $T_{1 \rightarrow 2}-T_{2 \rightarrow 1}\sim \frac{1}{2Cf}(\frac{V^2}{R_1}-p_{\rm c})=2.1$~K for $p_{\rm c}=\frac{V^2}{2}(\frac{1}{R_1}+\frac{1}{R_2})=46.8$~mW at which heating and cooling processes become symmetric~\cite{pc}. The essential point to cause periodic oscillation is discontinuity and hysteresis of resistance, which are characteristic of first-order transition. Since this simple model considers only two states with $R_1$ and $R_2$, it cannot precisely reproduce the observed waveform. Introducing temperature dependence of resistance and spatial variation of the phase of oscillation, we obtain a smooth variation of $I$ as shown by a dotted line in Fig.~4(b). Temperature dependence of resistance mainly comes from that in the metallic paths for a fixed conduction-path geometry. In the calculation of Fig.~4(b), we used $\frac{1}{R}\frac{{\rm d}R}{{\rm d}T}$~=~0.01~K$^{-1}$ deduced from the $R-T$ curve at $H=90$~kOe, where the whole sample is metallic.

Optimization of low-field magnetoresistance and current oscillation is our future goal. We believe that it will open new applications of the phase-separated states.
 
In conclusion, we observed anomalous transport properties in phase-separated manganites of (La$_{1-y}$Pr$_y$)$_{0.7}$Ca$_{0.3}$MnO$_3$ ($y$=0.7) and Nd$_{0.5}$Ca$_{0.5}$Mn$_{1-z}$Cr$_z$O$_3$ ($z$=0.03). Application of a large current causes a change from a metallic to an insulating state by Joule heating. This state supported by a large current is sensitive to magnetic fields and produces low field negative magnetoresistance effect. Application of a constant voltage causes oscillation of transport currents, which can be explained using a hysteretic and discontinuous transition between the metallic and insulating states.

\begin{acknowledgments}
This work is supported by Grant-in-aid for Scientific Research from the Ministry of Education, Culture, Sports, Science and Technology.
\end{acknowledgments}


\begin{thebibliography}{}
\bibitem{pc} As estimated before, $p_{\rm c}$ in the sample volume between voltage terminals is 9~mW. Since the distance between voltage terminals is about 1/5 of the sample length, 46.8~mW is a reasonable value for simulation. \label {pc}
\end{thebibliography}
\end{document}